# Effects of 3-d and 4-d-transition metal substitutional impurities on the electronic properties of $CrO_2$

M. E. Williams,[1] H. Sims,[2,3] D. Mazumdar,[2,3] W.H.Butler[2,3]

1. Department of Mathematics & Computer Science, University of Maryland Eastern Shore, Princess Anne, MD 21853

2. Center for Materials for Information Technology, University of Alabama, Tusacaloosa, AL 35401

3. Department of Physics, University of Alabama, University of Alabama, Tuscaloosa, AL 35401

## **Abstract**

We present first-principles based density functional theory calculations of the electronic and magnetic structure of $CrO_2$ with *3d* (Ti through Cu) and *4d* (Zr through Ag) substitutional impurities. We find that the half-metallicity of $CrO_2$ remains intact for all of the calculated substitutions. We also observe two periodic trends as a function of the number of valence electrons: if the substituted atom has six or fewer valence electrons (Ti-Cr or Zr-Mo), the number of down spin electrons associated with the impurity ion is zero, resulting in ferromagnetic (FM) alignment of the impurity magnetic moment with the magnetization of the $CrO_2$ host. For substituent atoms with eight to ten (Fe-Ni or Ru-Pd with the exception of Ni), the number of down spin electrons contributed by the impurity ion remains fixed at three as the number contributed to the majority increases from one to three resulting in antiferromagnetic (AFM) alignment between impurity moment and host magnetization. The origin of this variation is the grouping of the impurity states into 3 states with approximate "$t_{2g}$" symmetry and 2 states with approximate "$e_g$" symmetry. Ni is an exception to the rule because a Jahn-Teller-like distortion causes a splitting of the Ni $e_g$ states. For Mn and Tc, which have 8 valence electrons, the zero down spin and 3 down spin configurations are very close in energy. For Cu and Ag atoms, which have 11 valence electrons, the energy is minimized when the substituent ion contributes 5 down-spin electrons.  We find that the interatomic exchange interactions are reduced for all substitutions except for the case of Fe for which a modest enhancement is calculated for interactions along certain crystallographic directions.

## **Introduction**

Spintronic devices that make use of electron spin (in addition to charge) in order to control the flow of electrons have become increasingly important in recent years. These devices depend for their operation on spin-polarized currents, meaning that when a voltage is applied, unequal currents flow in the two spin-channels.  For simplicity, we ignore small spin-orbit coupling effects in this discussion.  The ultimate spin-polarized conductor is a half-metal; a material that is metallic in one spin channel, but insulating (or semiconducting) in the other because there is a gap in the density of states for that spin channel at the Fermi energy [1].

Several transition metal oxides, *e.g.* $Fe_3O_4$ [2, 3], $La_{2/3}Sr_{1/3}MnO_3$ (LSMO) [4], and $CrO_2$ [5-8] have been predicted to be half metals.  Experimentally, these systems have yielded high spin polarization values through direct and indirect measurement [9-15].  Rutile structure $CrO_2$ is a particularly interesting half-metal.  It is one of only a few ferromagnetic oxides (ferrimagnets are much more common) and is perhaps the only material that is expected to be a half-metal both in bulk and at its surface [16, 17, 18].  Experimentally, its spin polarization has been shown to be very close to 100%. [9,10]

Although the very high spin-polarization predicted by mean field electronic structure calculations has been experimentally confirmed, we should mention that the degree of electron correlation in $CrO_2$ remains a subject of debate and research [19-30].   The facts that $CrO_2$ is an excellent metal at low temperature [31,32], that its optical properties appear to be consistent with band theory predictions [23, 30] and that its relatively low magnetocrystalline anisotropy is more consistent with weak than strong correlations [24] have been used to argue for the validity of a weakly correlated mean field picture. On the other hand, the metal-insulator transition (which is only partially explained by band theory) in structurally and electronically similar $VO_2$ [33], and the results of recent calculations using dynamical mean field theory [29] have been used to argue for the importance of electron correlations beyond mean field. The calculations presented here are confined to the mean field electronic structure picture. One important prediction of our study is that early 4d substitutions generate electrons that are much more diffuse in both energy and spatial extent than those of the host. It is possible that such substitutions can be used to reduce the effects of any electron correlations.

Although the high spin-polarization and high conductivity of $CrO_2$ make it attractive for spintronics applications, its relatively low Curie temperature (~395K for bulk, somewhat lower for films) may cause problems for applications at room temperature and above. This work was initially motivated by a desire to investigate, theoretically, one possible way of altering and possibly enhancing the magnetic properties (*e.g.* Curie temperature) of $CrO_2$, namely through chemical doping [34]. Since Cr is a 3d-transition metal, we choose to substitute transition metal atoms from the 3d and 4d series for a Cr atom. Our calculations employ a supercell containing either four or eight formula units of $CrO_2$ (before substitution) making the new material either $Cr_{0.75}Z_{0.25}O_2$ or $Cr_{0.875}Z_{0.125}O_2$, respectively, where Z is any 3d (Ti, V, Mn, Fe, Co, Ni, or Cu) or 4d (Zr, Nb, Mo, Tc, Ru, Rh, Pd, or Ag) transition metal. Based on our calculations, we find, surprisingly, that none of the chemical dopants destroys the half-metal property of $CrO_2$ (although the minority band gap may be reduced). We also find that, although the 4*d* substitutions usually generate the same net change in magnetic moment as their 3*d* counterparts, they may do this in a somewhat different way; some of the moments induced by the 4d substituents being so diffuse as to negate the notion of "the moment on the impurity". Also surprisingly, we find that ferromagnetic metals such as Fe, Co, and Ni, favor anti-ferromagnetic alignment with the host Cr thereby lowering the cell moment and that non-magnetic elements tend to develop a magnetic moment when put into the host $CrO_2$ lattice. Detailed calculations of the exchange interaction were performed for the case of Fe substitution. The exchange interaction was found to be anisotropic and in certain directions we found an enhancement in exchange.

**Methodology**

We conducted first-principles based density function theory (DFT) calculations within the generalized gradient approximation (GGA) [35, 36] using the Vienna *ab-initio* Simulation package (VASP) [37, 38]. We constructed periodic supercells containing either 12 or 24 atoms based on multiples of the tetragonal 6-atom rutile cell having metal ions at the corners and body center. Details of the supercell shape are given in the section on exchange interactions. These cells allowed us to substitute either 25% or 12.5%, respectively, of the Cr ions with the chosen impurity ion. We used a 5 x 9 x 7 Monkhorst-Pack [MP] k-point grid. We allowed structural relaxation of the supercell but did not allow for the possibility of major defects in the rutile structure such as deviations from stoichiometry. The initial parameters of the cells were based on the known bulk $CrO_2$ values ($a=b=0.4593$nm, $c=0.2959$nm). Each system was relaxed so that the interatomic forces were less than 0.1 eV/Å, before the electronic structure and magnetic moments were calculated.

Plane wave based DFT codes generate charge and magnetization densities that extend throughout the computational cell. Any method of assigning a charge or magnetic moment to a particular ion is, to some extent, arbitrary. The total change in the magnetic moment on substituting an impurity atom for a Cr atom is, however, well and unambiguously defined. In these particular systems, moreover, we find that the minority Fermi energy continues to lie in a gap in the density of states even after the substitution. This implies that the number of minority electrons in the cell is an integer. Since the total number of electrons in the cell is also an integer, it follows that the number of majority electrons is an integer and finally that the total spin moment of the cell after the substitution is an integer. Thus, if we assign a moment of $2\mu_B$ to each of the Cr ions, we can associate the difference between the total cell moment after substitution and $2n\mu_B$ with the impurity moment. Here $n$ is the number of Cr ions in the cell after the substitution ($n=7$ for the calculations based on the 24 atom cell reported here). When we report magnetic moments for individual ions, they are calculated within a sphere of radius 0.9 Å for the 3$d$ ions and 1.0 Å for the 4$d$ ions. We do not discuss orbital moments in this paper.

## Results

### Electronic Structure of $CrO_2$

It is helpful to briefly review the DFT-GGA electronic structure of $CrO_2$ which we used as a starting point for understanding the effect of placing impurities into this host. The density of states (DOS) is shown in Figure 1 where the O-$p$ states and Cr-$d$ states are labeled. The assignment of states to Oxygen or Cr ions is, of course, approximate since all of the states are hybridized and the charge and magnetization densities from a DFT calculation cannot be assigned unambiguously to a particular ion. The number of states per atom, however, is largely conserved by the hybridization so that it is possible to describe the electronic structure in a manner that is consistent with a chemical picture.

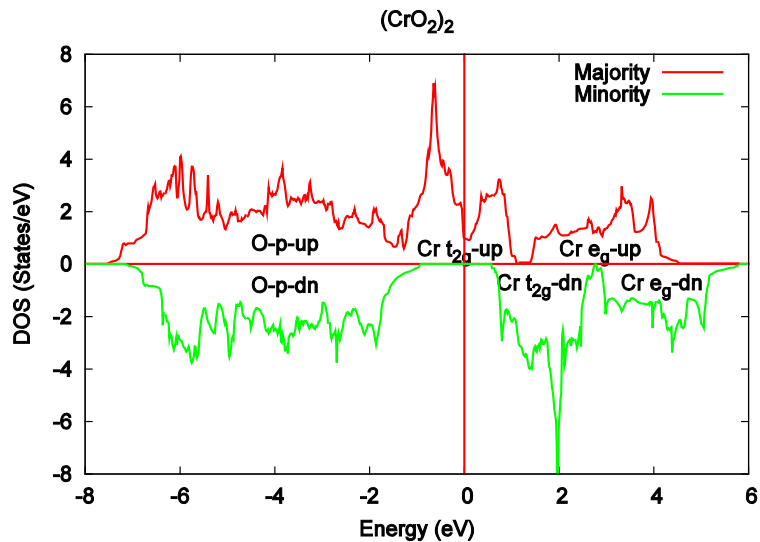

**Figure 1. Electronic Density of States of $CrO_2$.**

In this picture, the oxygen *p*-states are filled by each oxygen ion acquiring 2 additional electrons from the Cr ions. Since there are twice as many O ions as Cr ions, the O ions will be in the (nominal) -2 valence state and the Cr ions will be in the (nominal) +4 valence state. Each Cr ion is surrounded by 6 O ions which form a distorted octahedron. The approximate cubic crystal symmetry of the distorted octahedron breaks up the 5 Cr d-states per spin into a lower energy group of 3 per Cr ion that would be labeled "$t_{2g}$" if the symmetry were exactly cubic and a corresponding higher energy group of 2 "$e_g$" states per Cr ion. This is most easily seen for the minority Cr-d DOS where the DOS vanishes at an energy that separates a "lump" of DOS that contains 3 electrons from one that contains 2 electrons. For the majority channel, the strong exchange interactions push the majority "$t_{2g}$" states down in energy so that they overlap slightly with the majority O-p states. In a tight-binding model, this overlap between the O-p and Cr-d majority states requires some degree of direct Cr-*d* to Cr-*d* interaction.

The DFT-GGA DOS for non-magnetic $CrO_2$ is very similar to the magnetic version with the exception that the majority *d*-states are higher (identical to the minority) and well separated from the O-*p* states. The Fermi energy falls such that 1 majority and 1 minority Cr-*d* states are occupied. Thus in DFT, the system reduces its energy by generating spin-polarization that allows it to push 2 occupied Cr-d states down in energy. The system gains an additional reduction in energy because the majority Fermi energy falls in a pseudogap that forms within the "$t_{2g}$" complex.

## Cr(Ti)$O_2$ and Cr(Zr)$O_2$

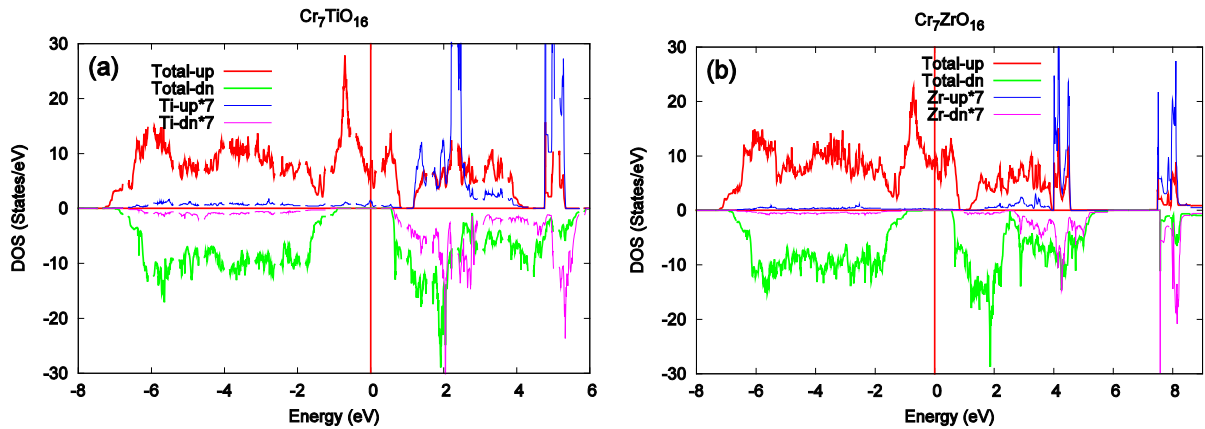

**Figure 2. (color on line) Calculated DOS for (a) Ti and (b) Zr impurities in $CrO_2$. The total DOS in the 24 atom supercell is indicated by thick lines. The DOS for the Ti and Zr impurities is indicated by thinner lines. The impurity DOS has been multiplied by 7 to enhance visibility. The impurity DOS is calculated within a sphere of radius 0.9 Å.**

Figure 2 shows the calculated DOS for Ti and Zr impurities in $CrO_2$. Since both impurity ions are surrounded by 6 oxygen ions, they are in a +4 valence state ($d^0$ configuration), the net moment on the Ti or Zr site is expected to be 0 so that each substituted Ti or Zr reduces the magnetic moment by $2\mu_B$. This is in qualitative agreement with our calculations. The net decrease in spin-moment associated with either substitution is exactly $2\mu_B$. This is clear from the DOS curves. Because the minority band gap is still present and the Fermi energy falls in it, the spin-moment is an integer and, because the number of occupied minority states does not change on substitution (the occupied minority states consist of the O 3-p states), all of the change in the number of electrons (-2 per substitution) must be accommodated in the majority channel. The moment within 0.9Å radius spheres surrounding the Ti or Zr sites is not exactly zero, but 0.105 for Ti and 0.03 for Zr.

One major difference between Ti and Zr is in the unoccupied part of the DOS where it can be seen that the Zr impurity $d$-states lie much higher in energy than those of Ti and have a larger splitting between the "$t_{2g}$" and "$e_g$" states. Another difference is that there is stronger hybridization of the Ti-$d$ states with the Oxygen-$p$ states due to their closer proximity in energy.

### $Cr(V)O_2$ and $Cr(Nb)O_2$

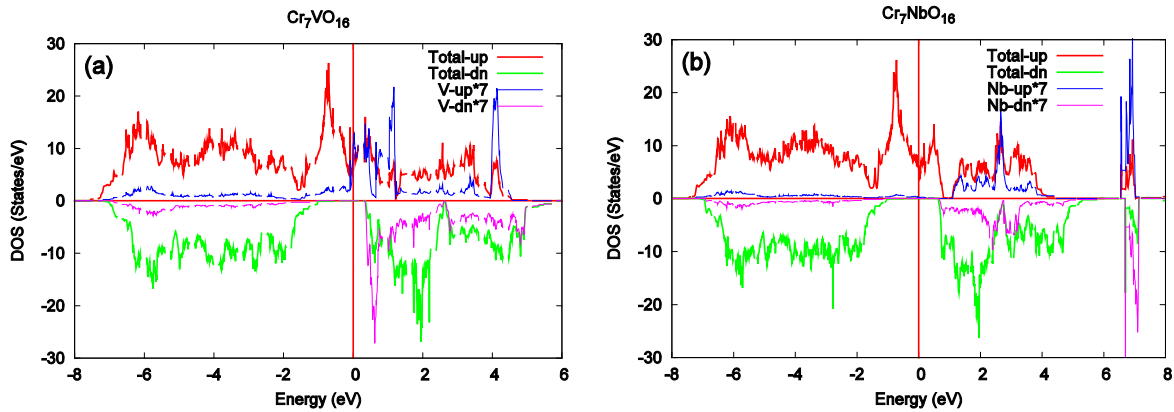

Figure 3. (Color online) Calculated DOS for (a) V and (b) Nb impurities in $CrO_2$. The total DOS in the 24 atom supercell is indicated by thick lines. The DOS for the V and Nb impurities is indicated by thinner lines and has been multiplied by 7. The impurity DOS is evaluated within a sphere of radius 0.9Å centered on the impurity site.

Figure 3 shows the calculated DOS for V and Nb impurities in $CrO_2$. Both of these impurity atoms have 5 valence electrons, one less than the Cr atom that they replace, so that when they are ionized by the quasi-octahedral oxygen environment, their configuration is expected to become $d^1$. We find that the total spin moment of the cell after the substitution is reduced from 16 to 15 indicating that the "V moment" and the "Nb moment" have aligned ferromagnetically with the magnetization of the host. The use of quotes in the preceding sentence is intended to indicate that the common notion of the spin moment "on" a transition metal ion may be an oversimplification, at least within this type of mean field calculation. For $CrO_2$, the moment within a 0.9Å radius Cr sphere is 1.796 or 90% of the total moment per Cr ion (with nominal spin $2\mu_B$). For the V impurity, the sphere moment is 0.455 $\mu_B$ or 45.5% of its nominal spin and for Nb it is only 0.069 $\mu_B$ or less than 7%.

We can obtain a qualitative understanding of the apparent missing magnetic moment from the information listed in Tables I and II. For the cell with the Nb substitution, the average spin moment within the 6 O ion spheres (radius 1.45Å) increases from $0.0105\mu_B$ to $0.0465\ \mu_B$. This would account for 0.22 $\mu_B$. The average moment within the 7 Cr spheres increases from 1.796 $\mu_B$ to 1.883 $\mu_B$. This is sufficient to account for 0.61 $\mu_B$. It should be noted that the sum of the sphere moments will not in general equal the total cell moment, but it is clear that the magnetization associated with the Nb substitution is quite diffuse.

The situation is similar for the V impurity, if less dramatic. In addition to the 0.455 $\mu_B$ within the V sphere, the average moment of the 6 neighboring O ions increases from 0.0105 to 0.0127 accounting for 0.013 $\mu_B$. The moment within the 7 Cr spheres increases from 1.796 $\mu_B$ to 1.851 $\mu_B$ accounting for 0.385 $\mu_B$.

### **Cr(Mo)O$_2$**

Figure 4 shows the calculated DOS for a Mo impurity in CrO2. Mo is iso-electronic with Cr. Both atoms have 6 valence electrons. Their ions in the octahedral O environment are expected to be in the $d^2$ state. The calculated magnetic moment of $16\mu_B$ for the $Cr_7MoO_{16}$ supercell is consistent with this state picture. However, from Figure 4 it appears that Mo is in something closer to a $d^0$ state since less than 12.5% of the expected $2\mu_B$ is within the 0.9Å radius sphere. This picture is also supported by the increased moment on the O ions surrounding the Mo impurity and by the enhancement of the moments within the Cr ion spheres as shown in Tables I and II.

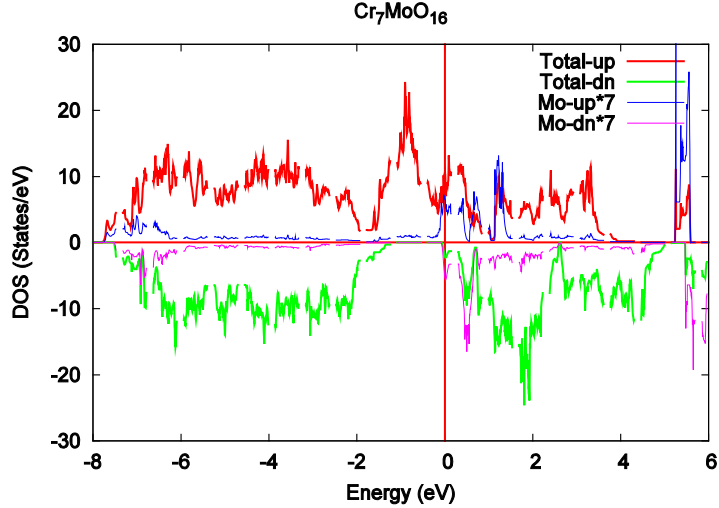

Figure 4. (Color online) Calculated DOS for Mo impurity in $CrO_2$. The total DOS in the 24 atom supercell is indicated by thick lines. The DOS for the Mo impurity site is indicated by thinner lines and has been multiplied by 7. The impurity DOS is evaluated within a sphere of radius 0.9Å centered on the impurity site.

From Figure 4, it is also clear that the shift of the impurity 4d-states to higher energy compared to 3d is reduced in Mo compared to Nb. The Mo minority impurity "$t_{2g}$" states lie just above the Fermi energy so that the system is only barely a half-metal. There is a large splitting between the "$t_{2g}$" and "$e_g$" impurity states.

## Cr(Mn)O$_2$ and Cr(Tc)O$_2$

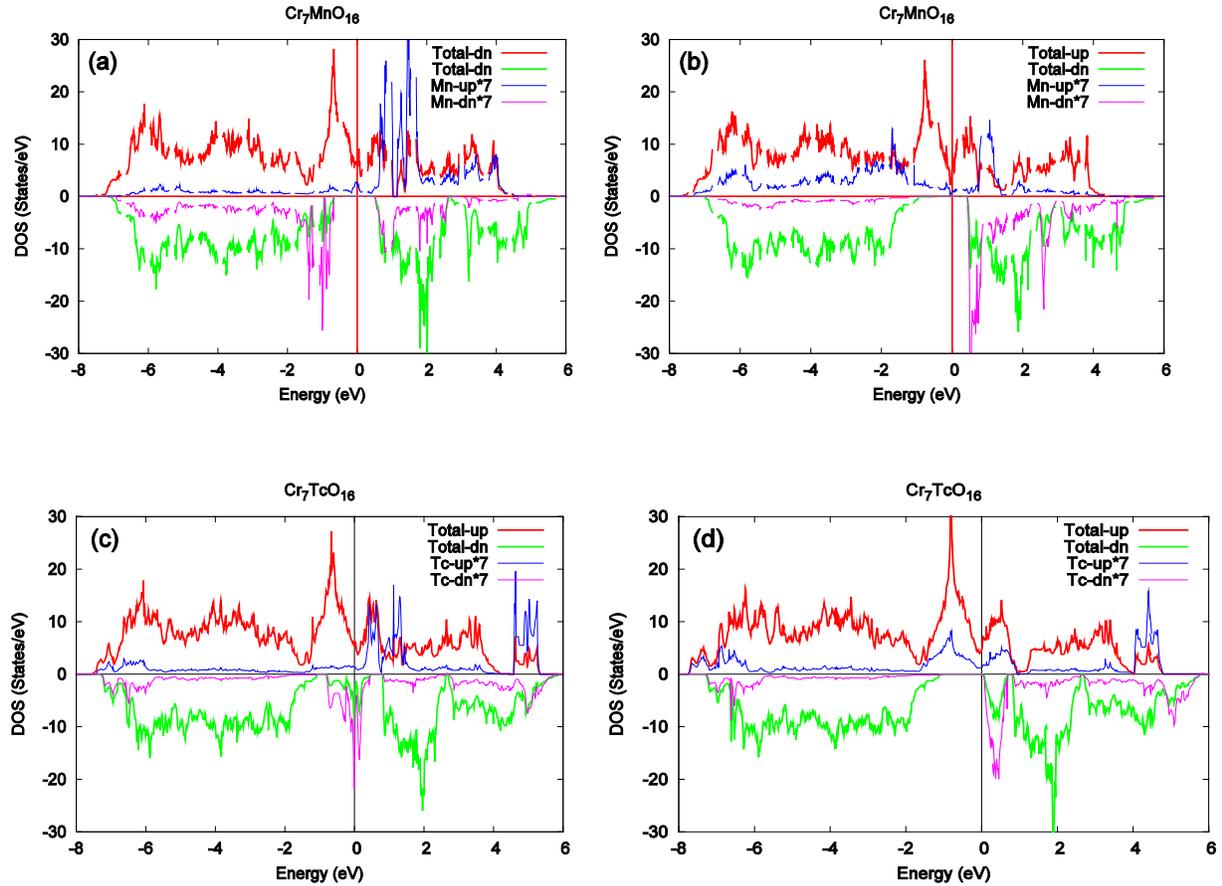

Figure 5. (Color online) Calculated DOS for Mn and Tc impurities in CrO$_2$. (a) Solution to DFT-GGA equations for Mn impurity with total moment per cell of 11$\mu_B$ and (b)17 $\mu_B$. (c) Solution to DFT-GGA equations for Tc impurity with total moment per cell of (c) 13.5 $\mu_B$ and (d) 17$\mu_B$. The total DOS in the 24 atom supercell is indicated by thick lines. The DOS for the Mn impurity site is indicated by thinner lines and has been multiplied by 7. The impurity DOS is evaluated within a sphere of radius 0.9Å centered on the impurity site.

Figure 5 shows the calculated DOS for a Mn impurity ((a) and (b)) and for a Tc impurity ((c) and (d)) in our 24 atom supercell. For all substitutions, we searched for both ferromagnetic (impurity moment aligned with host moments) and antiferromagnetic (impurity moment anti-aligned with host moments) solutions to the DFT-GGA equations. For the impurities presented in Figures 1-4, we were only able to obtain one solution for each impurity. For Mn and Tc impurities, however, we were able to obtain multiple solutions. For Mn, the two quite different solutions appear to be nearly degenerate in energy. One solution (panel a) generates a low moment. For this case we find a total cell moment of 11$\mu_B$, i.e. a net decrease of 5 $\mu_B$ on substitution of a Cr atom by a Mn atom. Since a Mn4+ ion is expected to be in the $d^3$ state, we would naturally associate the decrease of 5$\mu_B$ with a Mn ion having a moment of 3$\mu_B$ aligned antiferromagnetically. This is qualitatively what our calculation describes. The moment within the 0.9Å radius sphere is -2.17 $\mu_B$ rather than -3, but as we have emphasized the sphere radius is somewhat arbitrary. Part of the difference, is no doubt related to the strong hybridization of the minority Mn-d states with the O-p states that can be seen in Figure 5a. This picture is supported by the calculated moments within the O-ion spheres (radius=1.45Å) which average -0.026 $\mu_B$ on the 6 O-ions surrounding the Mn ion compared to +.0105 $\mu_B$ for the O spheres in $CrO_2$.

The other solution that we found (Figure 5b) gave a total cell moment of 17$\mu_B$ and was calculated to be approximately 3meV higher in energy than the 11 $\mu_B$ solution. The natural interpretation would be that in this case, the Mn $d^3$ ion with a spin moment of 3$\mu_B$ is aligned ferromagnetically with the host spin moments. This picture is consistent with the calculated moment within the (0.9Å radius) sphere of 2.52 $\mu_B$.

We also found two solutions for the Tc impurity. When the DFT-GGA equations are solved using VASP and following normal procedures beginning with a local impurity moment that is either aligned or anti-aligned one obtains a solution with a moment of approximately 13.5 $\mu_B$. For this solution (Figure 5c), the impurity "$t_{2g}$" states sit astride the Fermi energy and are approximately half-filled. We searched for additional states with 11 or 17 $\mu_B$ per cell corresponding to anti-ferromagnetic or ferromagnetic alignment of the impurity moment using a feature of VASP that allows one to constrain the total cell moment. Solutions with 11$\mu_B$ were not stable when the constraint was removed, but solutions with 17 $\mu_B$ were stable (Fig. 5d) and were lower in energy than the 13.5 $\mu_B$ solution (Fig 5d) by 173meV.

## Cr(Fe)O$_2$ and Cr(Ru)O$_2$

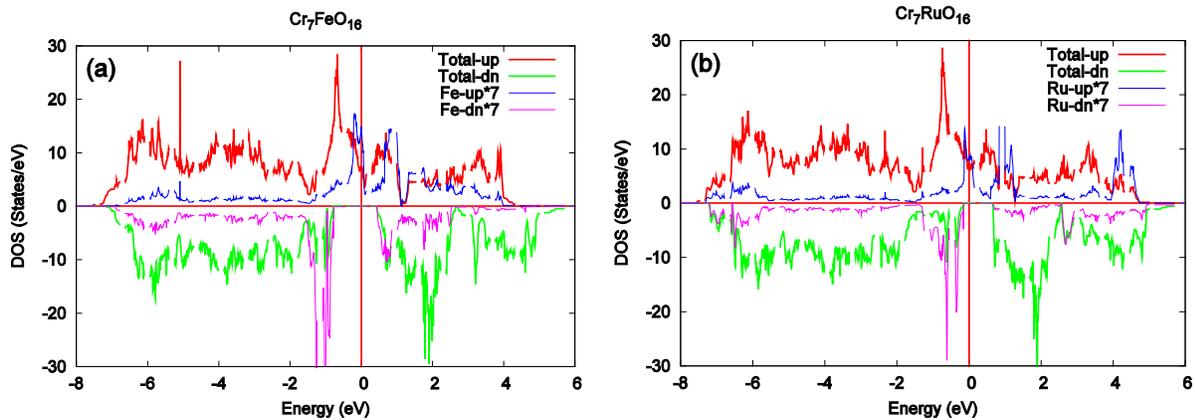

**Figure 6. . (Color online) Calculated DOS for (a) Fe and and (b) Ru impurities in CrO$_2$. The total DOS in the 24 atom supercell is indicated by thick lines. The DOS for the Fe and Ru impurities is indicated by thinner lines and has been multiplied by 7. The impurity DOS is evaluated within a sphere of radius 0.9Å centered on the impurity site.**

Figure 6 shows the DOS for (a) an Fe impurity and (b) a Ru impurity in CrO$_2$. Fe and Ru each have 8 valence electrons so their 4+ ions will have 4 electrons. The net change in moment due to the substitution is -4$\mu_B$ in both cases. From the DOS it is clear that this is accomplished by the three minority $t_{2g}$ states moving below the Fermi energy. For Fe, the impurity minority $t_{2g}$ states are moved down to the top of the O-p states so that the band gap is not greatly reduced from that of CrO$_2$. For Ru, on the other hand, the occupied minority $t_{2g}$ states are in the gap and significantly reduce it.

## Cr(Co)O$_2$ and Cr(Rh)O$_2$

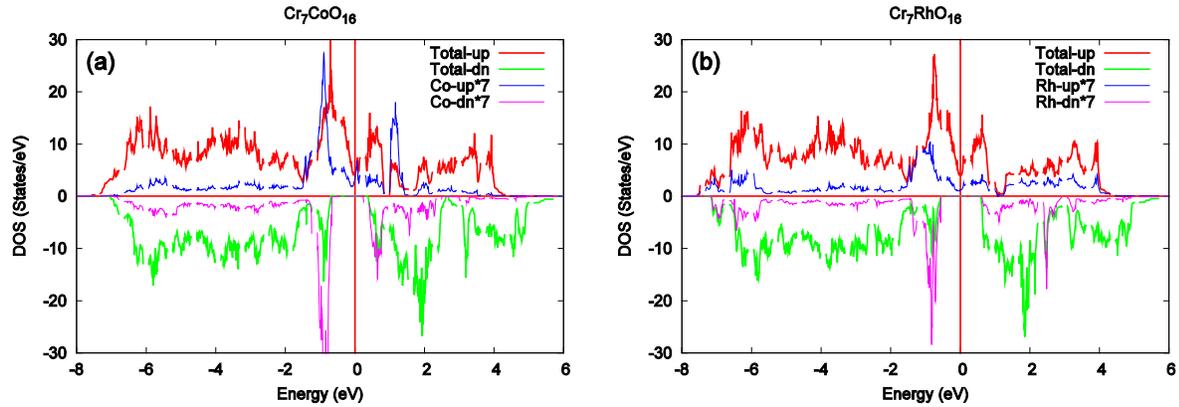

**Figure 7.** (Color online) Calculated DOS for (a) Co and and (b) Rh impurities in CrO$_2$. The total DOS in the 24 atom supercell is indicated by thick lines. The DOS for the CO and Rh impurities is indicated by thinner lines and has been multiplied by 7. The impurity DOS is evaluated within a sphere of radius 0.9Å centered on the impurity site.

Figure 7 shows the DOS for Co and Rh impurities in CrO$_2$. Co and Rh would have a d$^5$ configuration as 4+ ions. Similarly to Fe and Ru, for Co and Rh, the minority "t$_{2g}$" states are occupied. In both cases they are shifted down to the top of the minority O-p bands so that sizable minority band gaps survive the substitution.

## Cr(Ni)O$_2$ and Cr(Pd)O$_2$

Figure 8 shows the calculated DOS for Ni and Pd impurities in CrO$_2$. Ni and Pd impurities are expected to be in the d$^6$ configuration. If they follow the rule of filling the 3 minority "t$_{2g}$" states, these ions would contribute no magnetization to the cell so that the total moment would decrease by 2µ$_B$ per impurity. That is what happens for the Pd impurity. The three minority "t$_{2g}$" states are filled and the two minority "e$_g$" states are empty. The three minority states induced by the impurity must be compensated by the remaining three electrons contributed by the Pd ion so that the total moment induced by the impurity is simply -2µ$_B$, i.e. the moment that would have been contributed by the replaced Cr ion. For Ni, the case is somewhat different. The "$e_g$" levels are split with one "$e_g$" level being filled and the other empty. This leads to 4 minority electrons per impurity and 2 majority, so that the total moment induced by the impurity is -3$\mu_B$.

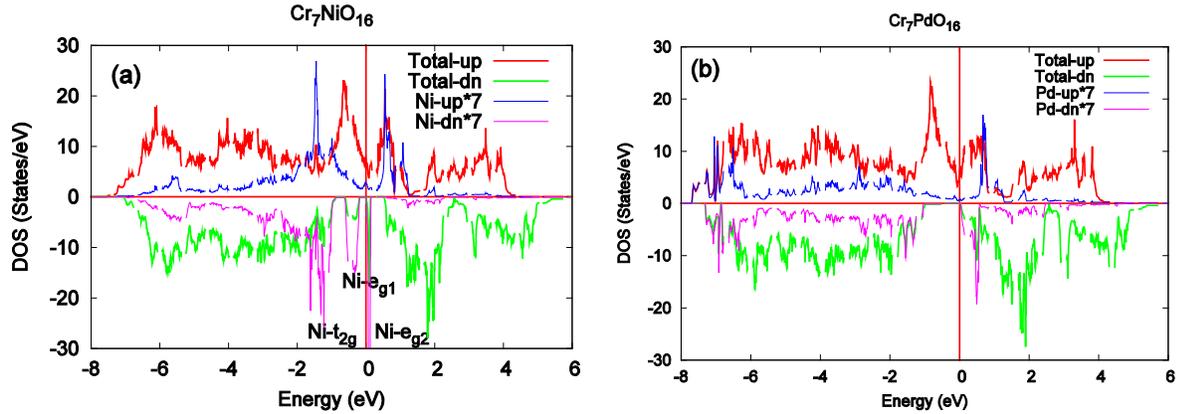

**Figure 8. (Color online) Calculated DOS for (a)Ni and and (b) Pd impurities in $CrO_2$. The total DOS in the 24 atom supercell is indicated by thick lines. The DOS for the Ni and Pd impurities is indicated by thinner lines and has been multiplied by 7. The impurity DOS is evaluated within a sphere of radius 0.9Å centered on the impurity site.**

Consider adding an additional electron to the DOS of Figure 7. One way that it can be done is exemplified by Pd, which simply adds it to the majority, pulling the majority states down slightly relative to the minority to accommodate the additional electron. The system with a Ni impurity adds the electron in a different way by distorting the octahedron so that the cation to apical O distances are increased from 1.91 Å to 1.99 Å. This causes the two "$e_g$" states to split. The system can then occupy the lower one and reduce its energy. This "strategy" works much better for Ni than for Pd because the Ni impurity "$e_g$" states have moved into the gap which makes them very narrow. It can be shown that because they are in the gap, they would be delta-functions if there were only interactions between the Cr ions and their surrounding O ions. In fact, there is significant direct interaction between neighboring Cr ions in the direction of the c axis. One of the "$e_g$" orbitals, however is oriented orthogonal to this direction so it is extremely narrow. The splitting of the "$e_g$" states in the CrO$_2$ gap is an example of a Jahn-Teller distortion [39].

## Cr(Cu)O$_2$ and Cr(Ag)O$_2$

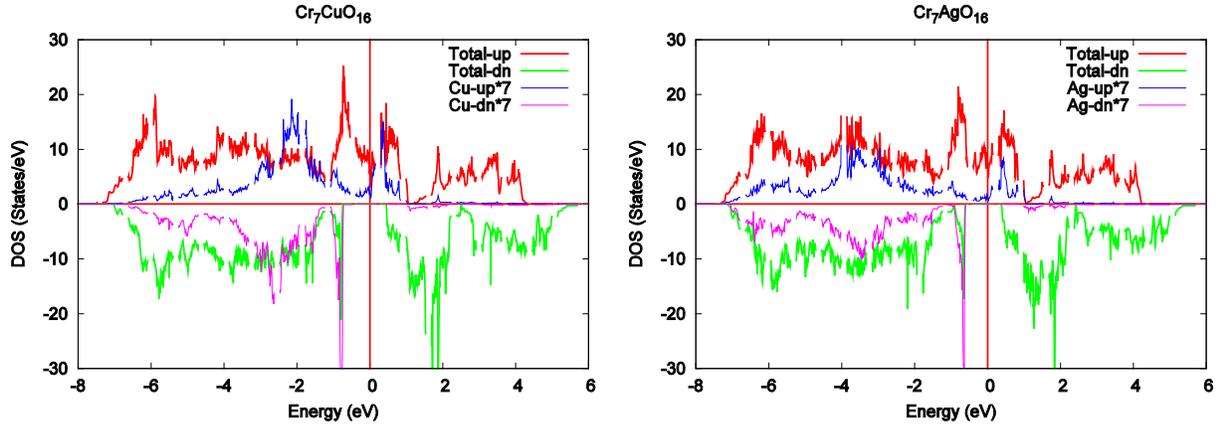

**Figure 9.** (Color online)  Calculated DOS for (a)Cu and (b) Ag impurities in CrO$_2$. The total DOS in the 24 atom supercell is indicated by thick lines. The DOS for the Cu and Ag impurities is indicated by thinner lines and has been multiplied by 7. The impurity DOS is evaluated within a sphere of radius 0.9Å centered on the impurity site.

The DOS for Cu and Ag impurities in CrO$_2$ is shown in Figure 9. Cu and Ag are expected to be in the $d^7$ configuration as impurities in CrO$_2$. From Figure 9, it is clear that they prefer to have all of their minority d-states occupied. In both cases the minority impurity $t_{2g}$ states hybridize strongly with the O-p states. The minority "$e_g$" states remain in the gap and occupy a very narrow energy range. Since five of the seven electrons of the $d^7$ ions go into the minority, the net change in magnetization induced by a Cu or Ag substitution is -5$\mu_B$, consisting of the loss of the two majority electrons "on" the Cr ion and the net three minority electrons "on" the Cu or Ag ions.

Table I: Summary of magnetic moment values for 3d substitution of $CrO_2$. Up (down) implies ferromagnetic (anti-ferromagnetic) alignment with host Cr. The magnetic moment units are $\mu_B$.

| Element | Ti | V | Cr | Mn | Fe | Co | Ni | Cu |
|---|---|---|---|---|---|---|---|---|
| **Nominal spin configuration of impurity** | 0 | 1 up | 2 up | 3 up or 3 down | 3 down & 1 up | 3 down & 2 up | 4 down & 2 up | 5 down & 2 up |
| **Total cell magnetic moment ($\mu_B$)** | 14 | 15 | 16 | 17 or 11 | 12 | 13 | 12 | 11 |
| **Impurity magnetic moment ($\mu_B$)** | 0.105 | 0.455 | n/a | 2.52 FM -2.169 AF | -1.269 | -0.167 | -0.515 | -0.599 |
| **Average Cr magnetic moment ($\mu_B$)** | 1.789 | 1.851 | 1.798 | 1.797 FM 1.759 AF | 1.766 | 1.728 | 1.660 | 1.591 |
| **Average magnetic moment of O ions ($\mu_B$)** | 0.002 | 0.009 | 0.010 | 0.023 FM -0.014 AF | -0.017 | -0.021 | -0.028 | -0.039 |
| **Average magnetic moment of 6 O ions nearest impurity ($\mu_B$)** | 0.020 | 0.016 | n/a | 0.051 FM -0.026 AF | -0.043 | -0.035 | -0.049 | -0.088 |
| **Distance to apical O ions (Å)** | 1.97 | 1.93 | 1.91 | 1.90 | 1.89 | 1.8☐ | 1.99 | 2.02 |
| **Distance to planar O ions (Å)** | 1.96 | 1.88 | 1.91 | 1.90 | 1.88 | 1.90 | 1.89 | 2.01 |

Table II: Summary of magnetic moment values for 4d substitution of $CrO_2$. Up (down) implies ferromagnetic (anti-ferromagnetic) alignment with host Cr. The magnetic moment units are $\mu_B$.

| Element | Zr | Nb | Mo | Tc | Ru | Rh | Pd | Ag |
|---|---|---|---|---|---|---|---|---|
| Nominal spin configuration of impurity | 0 | 1 up | 2 up | 3 up | 3 down & 1 up | 3 down & 2 up | 3 down & 3 up | 5 down & 2 up |
| Total cell magnetic moment ($\mu_B$) | 14 | 15 | 16 | 17 | 12 | 13 | 14 | 11 |
| Impurity magnetic moment ($\mu_B$) | 0.031 | 0.069 | 0.27 | 1.081 | -0.934 | -0.214 | 0.133 | -0.36 |
| Average Cr magnetic moment ($\mu_B$) | 1.786 | 1.883 | 1.942 | 1.904 | 1.768 | 1.748 | 1.767 | 1.563 |
| Average magnetic moment of O ions ($\mu_B$) | -0.007 | 0.015 | 0.015 | 0.029 | -0.042 | -0.027 | -0.0009 | -0.053 |
| Average magnetic moment of 6 O ions nearest impurity ($\mu_B$) | 0.028 | 0.047 | 0.045 | 0.082 | -0.074 | -0.040 | -0.020 | -0.094 |
| Distance to apical O ions (Å) | 2.09 | 2.01 | 1.96 | 1.95 | 1.97 | 1.98 | 2.00 | 2.15 |
| Distance to planar O ions (Å) | 2.09 | 1.99 | 1.94 | 1.96 | 1.98 | 2.00 | 2.02 | 2.16 |

## **Moment Rule**

Within DFT-GGA, the following rule appears to govern the ground state magnetic moment for 3d and 4d transition metal impurities in $CrO_2$: The number of minority electrons added by the impurity increases from zero (Ti, Zr, V, Nb, Cr, Mo, Mn, Tc) to three (Mn, Fe, Ru, Co, Rh, Ni[*], Pd) to five (Cu, Ag) as one proceeds across the 3d and 4d transition metal series. Mn is listed twice since the states with zero and three minority electrons added seem to be close in energy. Ni has an asterisk because it is an exception to the zero-three-five rule. The origin of this rule is almost certainly the grouping of states caused by the approximate cubic field of the oxygen ion distorted octahedron into a group of three and a group of two at higher energy coupled with the sharpness of these states as they cross the rather large gap in the minority DOS of $CrO_2$. Ni is the exception that helps to prove the rule since its ground state is accompanied by an exceptionally large distortion of the oxygen octahedron that causes a relatively large splitting of the two "$e_g$" states.

From the calculated DOS, the three "$t_{2g}$" and two "$e_g$" states are each spread over an energy range that varies from system to system, but is typically 2-3eV. Nevertheless, the calculated change in spin moment upon substitution is consistent with a picture in which these states have no dispersion. Thus the number of occupied minority d-states on the impurity is 0 ($d^0$-$d^3$ impurities) then jumps to 3 as all of the "$t_{2g}$" minority impurity states ($d^3$-$d^6$ impurities) are occupied and finally jumps to 5 as the two "$e_g$" minority impurity states are filled ($d^7$ impurities). Part of the reason for this behavior is that the dispersion of the minority impurity states may be significantly decreased if they fall in the minority gap of the $CrO_2$ where there are no d-states with which to interact. This result, at least for this system, helps to reconcile the band and crystal field pictures of transition metal oxides.

### Heisenberg Exchange Coupling in $Cr_{1-x}Z_xO_2$

Having determined the preferred magnetic alignment of the impurity ions within the $CrO_2$ lattice, we sought to investigate the change in the magnitude of the exchange coupling between Cr ions in the presence of the impurity ion. Within the Heisenberg model, the exchange energy is given by

$$H = -\sum_{i,j} J_{ij} S_i \cdot S_j$$

(1)

where $J_{ij}$ is the exchange energy between the $i$th and $j$th ions, $S_i$ is the unit vector pointing in the same direction as the moment on the $i$th ion, and we allow for the possibility that $i$ and $j$ are not limited to nearest neighbors. Therefore, if we assume that the magnitude of the Cr magnetic moments remains constant upon rotation, one can calculate the exchange energy, $J$, by comparing the total energy of a system in which all Cr ions are ferromagnetically aligned and one in which a given Cr ion (referred to hereafter as the "flipped Cr") is aligned antiferromagnetically to the rest (this difference in total energy will be referred to as the "total exchange energy"). In our calculations, the Cr moments decrease to about 80% of their proper magnitude when they are aligned antiferromagnetically, but one can simply scale the result to take this into account.

The primary contributions to the total exchange energy in $CrO_2$ are the interactions between neighbors along the c axis, those along the a or b axes, and between the corner and body-centered ions in the cell ($J_{001}$, $J_{100}$, and $J_{111}$, respectively). A thorough examination of the exchange interactions in bulk, impurity-free $CrO_2$ can be found in Ref. 40, in which it was found that $J_{001}$ and $J_{111}$ favor ferromagnetic alignment (with $|J_{001}| > |J_{111}|$), and $J_{100}$ favors antiferromagnetic alignment (but is smaller in magnitude than the others). Therefore, the position of the impurity ion with respect to the flipped Cr will be the factor that most affects the calculated total exchange energy.

We performed exchange calculations for the 3*d* impurity ions considered and found that, in almost every case, the total exchange energy for the impurity system was less than that of unmodified $CrO_2$. However, the Fe-doped system exhibited more complex behavior; the total exchange energy was enhanced in certain regions of the Cr-Fe-O system (for flipped Cr – Fe interactions along the a or b axes), reduced in others (for nearest neighbors along the c axis or the body diagonal), and nearly unaffected far from the impurity site. This dependence can be explained by returning to the calculations in which we determined the preferred orientation of the Fe ion in the Fe-doped system. There is also a dependence on the amount of Cr replaced by Fe.

In a simple one-unit-cell $Cr(Fe)O_2$ system (with one Cr and one Fe ion in the rutile cell), one can determine the magnitude of the Cr – Fe $J_{111}$, which is about half as strong as the same interaction in bulk $CrO_2$, by flipping the Fe ion and calculating the resulting change in energy. The total exchange energy of the calculation involving a flipped Cr neighboring the Fe impurity along the (111) direction will be made up of four Cr – Cr interactions and 4 Cr – Fe interactions each of which favors ferromagnetic alignment among the Cr ions. However, this energy will clearly be less than in the undoped system as the Cr – Fe interaction is weaker. We performed similar calculations to probe the Cr – Fe interactions along (100), (010) and (001), and found, unsurprisingly, that the interaction along the *a* and *b* axes was greater in magnitude than that in unmodified $CrO_2$, and the interaction along the *c* axis was smaller. Considering the average effect of the Fe impurity ion throughout the system, we find that, for 12.5% doping, the total exchange energy is decreased (by about 13%), as with the other impurity ions. For 25% doping,

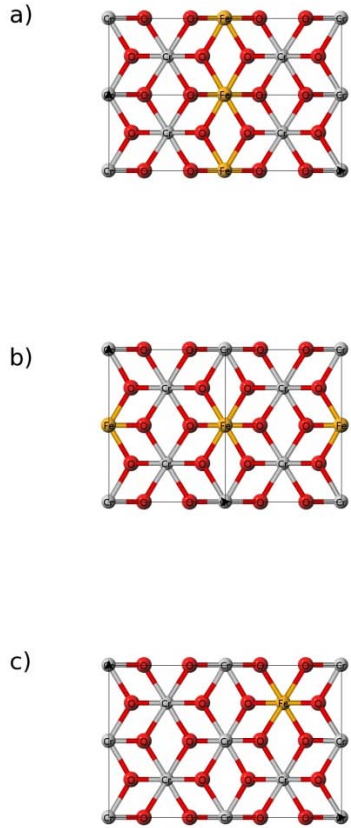

**Figure 10** a) Projection onto the *a-c* plane of the (a) 25% Fe-substituted cell with Fe ions occupying corner sites oriented along the *a* axis from the Cr ions. b) 25% Fe-substituted cell with Fe ions occupying corner sites oriented along the *c* axis from the Cr ions. c) 12.5% Fe-substituted cell. Shown is one possible placement of the Fe impurity in the 24-atom cell.

we considered two structures: one with the unit cells stacked in the (100) direction (as in Figure 10a) and one stacked in the (001) direction (Figure 10b). In the former case, the total exchange energy was decreased by about 16%, but in the latter it was increased by about 5%.

## **Conclusions**

We calculated the electronic structure $CrO_2$ with 3d and 4d transition metal ions as impurities. All of the considered systems were relaxed structurally, but we did not consider the possibility of major structural defects such as O ion vacancies. From our calculations, it can be seen that rutile $CrO_2$ remains a half-metal when impurity elements are substituted for Cr forming

$Cr_{0.875}Z_{0.125}O_2$ and $Cr_{0.75}Z_{0.25}O_2$. We find that the net change in spin magnetic moment can be understood in terms of a picture in which 0 ($d^0$-$d^3$ impurities), 3 ($d^3$-$d^6$ impurities) or 5 ($d^7$ impurities) electrons from the impurity become minority. This causes the net impurity moment to align antiferromagnetically with the Cr moments when the impurity has more than 3 valence electrons (the FM and AFM configurations are nearly degenerate in Mn). We found that impurity substitutions generally decreased the interatomic exchange interactions between the Cr ions compared to those in pure $CrO_2$. In the Fe-doped system, the exchange picture was more complex. As with the other impurities at 12.5% substitution, the magnitude and sign of the exchange energy depended on the location of the Fe impurity. Certain positions gave modest gains in the exchange energy, but on average the total exchange energy of the system was still decreased. At 25% substitution, the supercell seen in Figure (001fig) yielded a 5% increase in the total exchange energy. To further study these impurity systems, one should allow for structural defects beyond the small shifts in position considered here.

## **Acknowledgements**

The authors greatly appreciated input from Dr. Claudia Mewes of the MINT Center who read the paper and discussed various aspects with us. Some of the calculations were performed on the University of Alabama High-Performance Cluster. We also thank Professors Arunava Gupta and Patrick LeClair for helpful suggestions. This work was supported by the following grants: NSF MRSEC Grant No. 0213985 and NSF Rutiles project Grant No. 0706280.